\documentclass[12pt]{article}
\usepackage{mathtext}
\begin{document}
\title{\bf Quasar acceleration by a one-side jet and asymmetric radiation
      }
\author{
A.~I.~Tsygan
\\
{\small\it
Ioffe Physico-Technical Institute, }
\\
{\small\it
Politekhnicheskaya 26, 194021 Saint-Petersburg, Russia  
}
}
\date{} 

\maketitle

\begin{center}
{\large\bf Abstract }
\end{center}
A quasar (as well as an active galactic nucleus)
which emits a one-side jet (or which is an anisotropic
emitter of particles and quanta) can be accelerated and
leave its host galaxy.
\\ \

Let us consider a quasar or an active galactic nucleus
which emits a one-side jet or two antiparallel jets of different
intensities. The possibility of generating  one-side jets
has been discussed in some theoretical papers [1-3].
In this case an entire compact object in the center of
its host galaxy will be a subject of force opposite to
the jet direction, and the object will undergo acceleration.
The acceleration can also be produced by some asymmetry
of the quasar radiation. We will employ the standard
model in which the quasar is an accreting black
hole of mass $M_{\rm b}$. Let ${\rm d}M/{\rm d}t$ be the accretion
rate to the black hole from the disk which is fed up by stars
entering the region of tidal destruction. We denote the
mass of a compact stellar cluster by $M_{\rm s}$.

In the regime of the Eddington quasar luminosity
\begin{eqnarray}
   L_{\rm Edd}={4 \pi G M_{\rm b} m_p c \over \sigma_{\rm T}}
   \approx 1.25 \times 10^{46}  \,
   \left( M_{\rm b} \over 10^8\, M_\odot \right)\;\;{\rm erg~s}^{-1}
\nonumber
\end{eqnarray}
the accretion rate from the disk is determined by
$(\eta/0.1)\,c^2 \,{\rm d}M/{\rm d}t=L_{\rm Edd}$.
Here, $G$ is the gravitational constant,
$m_p$ is the proton mass, $c$ the velocity of light,
$\sigma_{\rm T}$ the Thomson scattering cross section, 
$M_\odot$ is the solar mass, and the factor $\eta$
describes the efficiency of the transformation of
the gravitational energy into the radiation energy.
The acceleration $a$ of the quasar is determined by
the equation of motion,
\begin{eqnarray}
  &&  (M_{\rm b}+M_{\rm s})\,a= {L \over v_0}= {\xi L_{\rm Edd} \over v_0},
\label{1} \\
 &&   a = \xi  { 4 \pi G m_p \over \sigma_{\rm T}}
       \left(  M_{\rm b} \over M_{\rm b}+M_{\rm s}  \right)\, {c \over v_0},
\nonumber
\end{eqnarray}
where $L=\xi\,L_{\rm Edd}$ is the jet-rocket power. The change of the
quasar mass during the quasar acceleration by the jet is neglected.

For a jet velocity $v_0 \sim c$ we obtain
\begin{equation}
         a \approx \xi   
         \left(  M_{\rm b} \over M_{\rm b}+M_{\rm s}  \right)
        \, 2 \times 10^{-6}\;\;{\rm cm~s}^{-2}.
\label{2}
\end{equation}
The factor $\xi$ describes the asymmetry of the quasar radiation
including the asymmetry of two jets. For a one-side
jet, $\xi$ is the fraction of energy which is carried away by
the jet. For certainty, we will use $\xi=0.1$
as a reference value. For
$M_{\rm b}=M_{\rm s}=10^8\,M_\odot$ we obtain the acceleration
$a \approx (\xi/0.1)\,10^{-7}$ cm~s$^{-2}$. In this case the accretion
rate is ${\rm d}M/{\rm d}t\approx 2.2\,(0.1/\eta)~~M_\odot$ yr$^{-1}$,
and the quasar life time is 
$\tau=M_{\rm s}/({\rm d}M/{\rm d}t)\approx (\eta /0.1)\,0.45\times
10^8$ yr.  At these parameters the quasar (the black hole, its
disk and the compact stellar cluster) can quietly leave
its host galaxy in $\tau \approx (\eta /0.1)\,0.45 \times 10^8$ yr
and fly away at a distance of $s=a \tau^2/2 \approx
32 (\xi/0.1)(\eta/0.1)^2$ kpc; its velocity will reach
$v=a \tau \approx 1.4 \times 10^8\,(\xi/0.1)(\eta/0.1)$ cm~s$^{-1}$.
During the entire acceleration stage the black hole will be fed up
by stars from its compact cluster of radius $r_{\rm s}$.
These stars are rather close to the black hole
(to satisfy the condition $a \ll GM_{\rm b}/r_{\rm s}^2$,
which is equivalent to $r_{\rm s} \ll \sqrt{GM_{\rm b}/a}
\approx 120\, \sqrt{0.1/\xi}$~pc) and follow it.

While estimating the quasar acceleration we have neglected its
interaction with the host galaxy. This is justified
provided
\begin{equation}
     G M(r)/r^2 \ll a,
\label{3}
\end{equation}
where $M(r)$ is the mass of a part of the galaxy
(excluding the quasar) confined in a sphere of radius $r$.
For $r=300$ pc the condition (\ref{3}) is fulfilled
at $M(r) \ll 6.5 \times 10^8 \,(\xi/0.1)\,M_\odot$
(and for $r=3$ kpc we would need $M(r) \ll 6.5 \times 10^{10}\,
(\xi/0.1)\,M_\odot$). If the condition (\ref{3}) is
violated at some radius $r_0$, the moving quasar will
be forced to drag the central part of its galaxy of the mass
$M(r_0) \sim a r_0^2/G$ which will reduce the quasar
acceleration.

Let us note that even if the jets are generated by the
electromagnetic mechanism [4], they may become asymmetric
if the magnetic field of the disk around the black hole
is odd with respect to the disk plane, $B_z(z)=-B_z(-z)$.
In this case one jet would contain accelerated electrons and the second
jet would contain accelerated protons. 
In the electron jet (in contrast to the
proton one) electron-positron pairs can be created,
which can screen an accelerating electric field and produce
some jet asymmetry.

Recently a bright quasar has been discovered outside a
host galaxy [5]. It has been suggested to be thrown away
during merging of two galaxies one of which had a binary
nucleus [6]. Our alternative explanation is that the quasar
left its host galaxy under the proposed rocket effect.

\section*{Acknowledgments}

This work was partly
supported by the
Russian Foundation for Basic Research (grant 04-02-17590),
and by the Russian Leading Scientific School Program
(grant 9879.2006.2).


\begin{thebibliography}{99}
\bibitem{Wiita}
Wiita, P.J., Siah, M.J., 1981, ApJ., {\bf 243}, 710
\bibitem{Icke}
Icke, V., 1983, ApJ., {\bf 265}, 648
\bibitem{Bodo}
Bodo, G., Chagelishvili, G.D., Ferrari, A., Lominadze, J.G., 
Trussoni, E., Plasma Astrophysics, June 1990, Telavi, Georgia, USSR, p. 273.
\bibitem{Blandford}
Blandford, R.D., Znajek R.L., 1977, MNRAS, {\bf 179}, 433
\bibitem{Magain}
Magain, P., Letawe, G., Courbin, F., Jablonka, P., Jahnke, K.,
Meylan, G., 2005, Nature, {\bf 437}, 381
\bibitem{Hoffman}
Hoffman, L., Loeb, A., astro-ph/0511242 v1 8 Nov 2005
\end{thebibliography}
\end{document}